\documentclass[prd,nofootinbib,onecolumn,preprint,11pt]{revtex4}

\usepackage{graphicx}
\usepackage{bm}
\usepackage{float}
\usepackage{subfigure}
\usepackage{amsmath}
\usepackage{amsfonts}
\usepackage{color}
\usepackage{slashed}
\usepackage{fancyhdr}

\newcommand{\gb}{\bar{g}}

\newcommand{\sbar}{\bar{\sigma}}

\newcommand{\dx}[1]{\text{d}#1}
\newcommand{\dd}{\text{d}}

\newcommand{\D}{\mathcal{D}}

\begin{document}

\title{A Line Source In Minkowski For The de Sitter Spacetime Scalar Green's Function: Massless Minimally Coupled Case}

\author{Yi-Zen Chu}
\affiliation{
Center for Particle Cosmology, Department of Physics and Astronomy, University of Pennsylvania, Philadelphia, Pennsylvania 19104, USA
}

\begin{abstract}
\noindent Motivated by the desire to understand the causal structure of physical signals produced in curved spacetimes -- particularly around black holes -- we show how, for certain classes of geometries, one might obtain its retarded or advanced minimally coupled massless scalar Green's function by using the corresponding Green's functions in the higher dimensional Minkowski spacetime where it is embedded. Analogous statements hold for certain classes of curved Riemannian spaces, with positive definite metrics, which may be embedded in higher dimensional Euclidean spaces. The general formula is applied to $(d \geq 2)$-dimensional de Sitter spacetime, and the scalar Green's function is demonstrated to be sourced by a line emanating infinitesimally close to the origin of the ambient $(d+1)$-dimensional Minkowski spacetime and piercing orthogonally through the de Sitter hyperboloids of all finite sizes. This method does not require solving the de Sitter wave equation directly. Only the zero mode solution to an ordinary differential equation, the ``wave equation" perpendicular to the hyperboloid -- followed by a one dimensional integral -- needs to be evaluated. A topological obstruction to the general construction is also discussed by utilizing it to derive a generalized Green's function of the Laplacian on the $(d \geq 2)$-dimensional sphere.
\end{abstract}

\maketitle

\section{Motivation}

In about 2 decades or so, humanity may have the capability, through a space based gravitational wave detector such as eLISA/NGO, to listen to the spacetime ripples generated by compact bodies orbiting around and plunging into the super-massive black holes residing at the center of galaxies. To model the dynamics of these compact bodies, it is necessary to have a quantitative understanding of the causal structure of the gravitational field generated by them. This is because gravitons, despite their massless nature in Minkowski spacetime, travel both on and inside the forward light cone of their sources in generic curved spacetimes \cite{Nomenclature}. As a result, the trajectory of these compact bodies at any given point in time depend, in principle, on their entire past histories.

In this paper, we wish to initiate a program of understanding the causal structure of physical signals in curved geometries by embedding these geometries in an appropriate higher dimensional Minkowski spacetime. The main reason for doing so is because the 4-dimensional (4D) Schwarzschild black hole spacetime has explicitly known embeddings in 6D Minkowski -- see, for instance, Fronsdal \cite{Fronsdal:1959zza}. At this point it is uncertain if this program will be successful for the Schwarzschild black hole. But we wish to show, for $(d \geq 2)$-dimensional de Sitter, which can be viewed as a hyperboloid in $(d+1)$-dimensional flat spacetime, the causal structure of minimally coupled massless scalar waves in the former can be readily illuminated in terms of massless scalar waves in the latter. Specifically we will discuss in some detail how the retarded or advanced Green's function of the massless scalar in $d$-dimensional de Sitter is sourced by a line, in the $(d+1)$-dimensional flat spacetime, starting infinitesimally close to the Minkowski origin and running perpendicularly through the de Sitter hyperboloid. The spacetime location where this line source intersects the hyperboloid is interpreted by the observer (living in de Sitter spacetime) as the spacetime point source of the Green's function \cite{NoteAdded}.

In section \eqref{Section_Generalities}, we will derive the general construction, leading up to the formula in eq. \eqref{GreensFunction_GeneralFormula}. Following that we will review the Green's functions of the Minkowski wave operator and the Euclidean Laplacian in all dimensions, highlighting how one may use the embedding viewpoint to obtain them from the $d=3$ Minkowski case. In section \eqref{Section_deSitter} we will then apply the method to obtain the de Sitter spacetime minimally coupled massless scalar Green's function, and in section \eqref{Section_nSphere} we discuss a topological obstruction to the general construction. There, we will propose a generalized Green's function for the Laplacian on the $d$-sphere. We summarize our findings and lay out possible future directions in section \eqref{Section_FutureDirections}.

\section{Generalities}
\label{Section_Generalities}

We shall first begin in $(d+n)$-dimensional Minkowski spacetime endowed with Cartesian coordinates $\{X^\mathfrak{A} | \mathfrak{A} = 0,1,2,\dots,d+n-1 \}$. Its metric is
\begin{align}
\label{MinkowskiMetric}
\dd s^2 						&= \eta_{\mathfrak{A}\mathfrak{B}} \dd X^{\mathfrak{A}} \dd X^{\mathfrak{B}} \equiv \dd X \cdot \dd X , \\
\eta_{\mathfrak{A}\mathfrak{B}} &\equiv \text{diag}[1,-1,\dots,-1].
\end{align}
Suppose we found a set of coordinates $\{x^\mu,y^\text{A}|\mu=0,1,2,\dots,d-1; \ \text{A} = 1,2,3,\dots,n \}$ that transforms the $(d+n)$-dimensional Minkowski metric in \eqref{MinkowskiMetric}, via
\begin{align}
X^{\mathfrak{A}} 
	&\to X^{\mathfrak{A}}[x,y] \\
\dd X^{\mathfrak{A}} 
	&\to \frac{\partial X^{\mathfrak{A}}[x,y]}{\partial x^\alpha} \dd x^\alpha + \frac{\partial X^{\mathfrak{A}}[x,y]}{\partial y^\text{B}} \dd y^\text{B},
\end{align}
into the form
\begin{align}
\label{Metric_GeneralEmbeddingForm}
\dd s^2 &= P^2[y] \bar{g}_{\mu\nu}[x] \dd x^\mu \dd x^\nu + g^\perp_\text{AB}[y] \dd y^\text{A} \dd y^\text{B} ,
\end{align}
where $\bar{g}_{\mu\nu}$ does not depend on $y$ while $P$ and $g^\perp_\text{AB}$ do not depend on $x$. Suppose further that the curved spacetime metric $g_{\mu\nu}[x]$ of interest is realized on the surface $y=y_0$, i.e.
\begin{align}
\label{Metric_OfActualInterest}
g_{\mu\nu}[x] = P^2[y_0] \bar{g}_{\mu\nu}[x] ,
\end{align}
with its wave operator $\Box$ acting on a scalar $\psi$ as
\begin{align}
\label{ScalarWaveOperator_gmunu}
\Box \psi 
= \frac{1}{P^2[y_0] \sqrt{|\gb|}} \partial_\mu \left( \sqrt{|\gb|} \gb^{\mu\nu} \partial_\nu \psi \right) 
\equiv \frac{1}{P^2[y_0]} \overline{\Box} \psi ,
\end{align}
where $\sqrt{|\gb|}$ is the square root of the absolute value of the determinant of $\gb_{\mu\nu}$ and $\gb^{\mu\nu}$ is its inverse. Then the Green's function $G_d[x,x']$ -- whose solution is the central goal of this paper -- obeying the equation
\begin{align}
\label{GreensFunction_DefiningEquation}
\Box_x G_d[x,x'] = \Box_{x'} G_d[x,x'] 
= \frac{\delta^{(d)}[x-x']}{\left\vert g[x] g[x'] \right\vert^{1/4}} 
= \frac{\delta^{(d)}[x-x']}{P^d[y_0] \left\vert \gb[x] \gb[x'] \right\vert^{1/4}} 
\end{align}
can be deduced from its $(d+n)$-dimensional flat spacetime counterpart $\overline{G}_{d+n}[X-X']$, through the general formula
\begin{align}
\label{GreensFunction_GeneralFormula}
G_d[x,x'] 
= \int \dd^n y' \sqrt{|g^\perp[y']|} \frac{P^{d-2}[y']}{P^{d-2}[y_0]} 
 \frac{\psi_{\sigma_{(0^\perp)}}^\dagger[y']}{\psi_{\sigma_{(0^\perp)}}^\dagger[y_0]} \overline{G}_{d+n}\Big[ X[x,y_0] - X'[x',y'] \Big] .
\end{align}
The $\overline{G}_{d+n}[X-X']$ themselves, which can be found in equations \eqref{GreensFunction_Minkowski_RetardedAdvanced} through \eqref{GreensFunction_Minkowski_Odd} below, obey the equations
\begin{align}
\label{GreensFunction_Minkowski_CartesianWaveEquation}
\eta^{\mathfrak{A}\mathfrak{B}} \partial_{\mathfrak{A}}\partial_{\mathfrak{B}} \overline{G}_{d+n}[X-X']
&= \eta^{\mathfrak{A}\mathfrak{B}} \partial_{\mathfrak{A}'}\partial_{\mathfrak{B}'} \overline{G}_{d+n}[X-X'] \nonumber\\
&= \delta^{(d+n)}[X-X'] = \frac{\delta^{(d)}[x-x']}{\sqrt{P^d[y] P^d[y']} \sqrt[4]{\gb[x] \gb[x']}} \frac{\delta^{(n)}[y-y']}{\sqrt[4]{g^\perp[y] g^\perp[y']}},
\end{align}
with the unprimed indices denoting derivatives with respect to $X$ and the primed indices derivatives with respect to $X'$. The $g^\perp[y']$ is the determinant of the metric $g^\perp_\text{AB}[y']$; the $\psi_{\sigma_{(0^\perp)}}$ is any one of the solutions to the zero mode equation
\begin{align}
\label{ZeroModeEquation}
\D_y \psi_{\sigma_{(0^\perp)}}[y] 
\equiv \frac{\partial_\text{A} \left( P^d \sqrt{|g^\perp|} (g^\perp)^\text{AB} \partial_\text{B} \psi_{\sigma_{(0^\perp)}} \right)}{P^{d-2} \sqrt{|g^\perp|}}
&= 0 ;
\end{align}
and $(g^\perp)^\text{AB}$ is the inverse of $g^\perp_\text{AB}$. 

{\bf Justification of eq. \eqref{GreensFunction_GeneralFormula}} \qquad Consider applying $\overline{\Box}_{x'}$ to both sides of the general formula in eq. \eqref{GreensFunction_GeneralFormula}.
\begin{align}
\label{GreensFunction_GeneralFormula_Justification_I}
\overline{\Box}_{x'} G_d[x,x'] 
= \int \dd^n y' \sqrt{|g^\perp[y']|} \frac{P^{d-2}[y']}{P^{d-2}[y_0]} 
 \frac{\psi_{\sigma_{(0^\perp)}}^\dagger[y']}{\psi_{\sigma_{(0^\perp)}}^\dagger[y_0]} \overline{\Box}_{x'} \overline{G}_{d+n}\Big[ X[x,y_0] - X'[x',y'] \Big] .
\end{align}
Now, a direct calculation would reveal that, for any scalar function $\psi$,
\begin{align}
\label{ScalarWaveOperator_SeparatedForm}
\overline{\Box}_{x'} \psi = P^2[y'] \eta^{\mathfrak{A'}\mathfrak{B'}} \partial_{\mathfrak{A}'} \partial_{\mathfrak{B}'} \psi - \D_{y'} \psi .
\end{align}
Because of eq. \eqref{GreensFunction_Minkowski_CartesianWaveEquation}, we may use eq. \eqref{ScalarWaveOperator_SeparatedForm} in eq. \eqref{GreensFunction_GeneralFormula_Justification_I}.
\begin{align}
\label{GreensFunction_GeneralFormula_Justification_II}
&\overline{\Box}_{x'} G_d[x,x'] 
= \frac{\delta^{(d)}[x-x']}{P^{d-2}[y_0] \sqrt[4]{\gb[x] \gb[x']}} \\
&\qquad\qquad
- \int \dd^n y' \sqrt{|g^\perp[y']|} \frac{P^{d-2}[y']}{P^{d-2}[y_0]} 
\frac{\D_{y'} \psi_{\sigma_{(0^\perp)}}^\dagger[y']}{\psi_{\sigma_{(0^\perp)}}^\dagger[y_0]} \overline{G}_{d+n}\Big[ X[x,y_0] - X'[x',y'] \Big] .
	\nonumber
\end{align}
Here, we have integrated-by-parts the operator $\D_{y'}$ and shifted it to act on $\psi_{\sigma_{(0^\perp)}}^\dagger[y']$. This assumes that the boundary terms are zero, which can be checked explicitly for the main results in this paper. Also, in eq. \eqref{GreensFunction_GeneralFormula_Justification_II}, we see the reason for the $P^{d-2}[y_0] \psi_{\sigma_{(0^\perp)}}^\dagger[y_0]$ in the denominator: this ensures the measure multiplying the $\delta$-functions on the right hand side is the desired one.

We see, therefore, that the Green's function equation eq. \eqref{GreensFunction_DefiningEquation} is satisfied if $\psi_{\sigma_{(0^\perp)}}$ obeys the zero mode equation \eqref{ZeroModeEquation}.

{\bf A check} \qquad Let us now apply $\overline{\Box}_x$ instead of $\overline{\Box}_{x'}$ on both sides of the general formula in eq. \eqref{GreensFunction_GeneralFormula},
\begin{align}
\overline{\Box}_x G_d[x,x'] 
= \int \dd^n y' \sqrt{|g^\perp[y']|} \frac{P^{d-2}[y']}{P^{d-2}[y]} 
\frac{\psi_{\sigma_{(0^\perp)}}^\dagger[y']}{\psi_{\sigma_{(0^\perp)}}^\dagger[y]} 
\overline{\Box}_x \overline{G}_{d+n}\Big[ X[x,y] - X'[x',y'] \Big] .
\end{align}
For now, we have left the $y$ unevaluated, i.e., replaced $y_0$ with $y$. Let us re-express $\overline{\Box}$ in terms of $\eta^{\mathfrak{A}\mathfrak{B}} \partial_\mathfrak{A} \partial_\mathfrak{B}$ and $\D$, using eq. \eqref{ScalarWaveOperator_SeparatedForm}. Upon recalling eq. \eqref{GreensFunction_Minkowski_CartesianWaveEquation},
\begin{align}
\label{GreensFunction_GeneralFormula_Check_IofII}
&\overline{\Box}_x G_d[x,x'] 
= \frac{\delta^{(d)}[x-x']}{P^{d-2}[y] \left\vert \gb[x'] \gb[x'] \right\vert^{1/4}} \\
&\qquad
- \frac{1}{P^{d-2}[y]} \frac{1}{\psi_{(\sigma_{(0^\perp)})}^\dagger[y]} \D_y 
\left(\int \dd^n y' \sqrt{|g^\perp[y']|} P^{d-2}[y'] \psi_{(\sigma_{(0^\perp)})}^\dagger[y'] \overline{G}_{d+n}\Big[ X[x,y]-X'[x',y'] \Big]\right). \nonumber
\end{align}
If the $\psi_{(\sigma_{(0^\perp)})}^\dagger[y] =$ constant solution were used, as it will be for the cases considered in this paper, eq. \eqref{GreensFunction_GeneralFormula_Check_IofII} simplifies to
\begin{align}
\label{GreensFunction_GeneralFormula_Check_IIofII}
\frac{1}{P^2[y]} \overline{\Box}_x G_{d}[x,x'] 
&= \frac{\delta^{(d)}[x-x']}{P^d[y] \left\vert \gb[x'] \gb[x'] \right\vert^{1/4}} - \frac{1}{P^d[y]} \D_y \mathcal{G}_d
\end{align}
with
\begin{align}
\label{GreensFunction_GeneralFormula_Check_Integral}
\mathcal{G}_d \equiv \int \dd^n y' \sqrt{|g^\perp[y']|} P^{d-2}[y'] \overline{G}_{d+n}\Big[ X[x,y]-X'[x',y'] \Big] .
\end{align}
When $y=y_0$, eq. \eqref{GreensFunction_GeneralFormula_Check_IIofII} is the same wave equation in eq. \eqref{GreensFunction_DefiningEquation} if and only if the integral $\mathcal{G}_d$ is itself a zero mode solution of eq. \eqref{ZeroModeEquation}, i.e., iff it is annihilated by the $\D$. Therefore, for a given metric and its embedding in Minkowski spacetime, one need only to write down the corresponding $\mathcal{G}_d$ in eq. \eqref{GreensFunction_GeneralFormula_Check_Integral}. If the coordinate(s) $y$ can be re-scaled away before the integral is even evaluated, so that $\mathcal{G}_d$ becomes independent of $y$, then the solution of the Green's function wave equation \eqref{GreensFunction_DefiningEquation} is assured. We will, in fact, use these observations in the forthcoming sections as a check that our calculations are solving the proper wave equations.

{\bf Causal structure} \qquad We have just seen that the Green's function in the $d$ dimensional curved spacetime, given in eq. \eqref{GreensFunction_GeneralFormula}, is the flat spacetime Green's function in $(d+n)$ dimensions projected along the zero modes of $\D$ in eq. \eqref{ZeroModeEquation}. More geometrically, however, we can read eq. \eqref{GreensFunction_GeneralFormula} as saying that the observer on the curved spacetime is really getting bathed by scalar particles produced from an (appropriately weighted) $n$-dimensional membrane in the ambient Minkowski, intersecting perpendicularly with the observer's world.

Analogous statements hold, if we replace in this section all instances of ``Minkowski" and ``flat spacetime" with ``Euclidean space," $\eta_{\mathfrak{A}\mathfrak{B}}$ with $\delta_{\mathfrak{A}\mathfrak{B}}$ (the Kronecker delta), and ``curved spacetime" for ``curved Riemannian spaces" (with positive definite metrics). However, a subtlety that is present for curved spacetimes embedded in Minkowski but not for that of curved spaces in Euclidean, is the notation of casual influence: when cause precede effect (or effect precede cause) in the ambient Minkowski spacetime, under what circumstances do cause precede effect (effect precede cause) in the curved spacetime itself? In the same vein -- when observer and source are within each other's light cone from the Minkowski perspective, does the same hold from the curved spacetime viewpoint too? We merely raise these issues here in the abstract, but will address it in some detail for de Sitter spacetime in section \eqref{Section_deSitter}.

\section{Flat Space(time)s}
\label{Section_Flat}

Because the flat space(time) Green's functions lie at the core of the embedding calculations below, in this section, we record their explicit expressions in all relevant dimensions $d$. Both the Minkowski Green's functions and their Euclidean counterparts in various dimensions are intimately connected to each other \cite{SoodakTiersten} -- we will illuminate this using the embedding viewpoint and eq. \eqref{GreensFunction_GeneralFormula}.

{\bf Minkowski} \qquad First define
\begin{align}
\label{Minkowski_SyngeWorldFunction}
\bar{\sigma} \equiv \frac{1}{2} \left(X-X'\right)^2 
\equiv \frac{1}{2} \eta_{\mathfrak{A}\mathfrak{B}} \left(X-X'\right)^\mathfrak{A} \left(X-X'\right)^\mathfrak{B} .
\end{align}
$\bar{\sigma}$, known as Synge's world function, is half the square of the geodesic distance between $X$ and $X'$ in the $(d+n)$-dimensional flat spacetime. Also define the step function
\begin{align}
\Theta[z] 
&= 1, \qquad z \geq 0 \nonumber\\
&= 0, \qquad z < 0 .
\end{align}
The retarded $\overline{G}^+_d[X-X']$ and advanced $\overline{G}^-_d[X-X']$ Green's functions are
\begin{align}
\label{GreensFunction_Minkowski_RetardedAdvanced}
\overline{G}^\pm_d[X-X'] = \Theta[\pm(X^0-X'^0)] \bar{\mathcal{G}}_d[\sbar] .
\end{align}
For even $d \geq 2$, the symmetric Green's function $\bar{\mathcal{G}}_d[\sbar]$ is \cite{GConfusion}
\begin{align}
\label{GreensFunction_Minkowski_Even}
\bar{\mathcal{G}}_{\text{even }d} \left[\sbar\right] 
= \frac{1}{2(2\pi)^{\frac{d-2}{2}}} \left(\frac{\partial}{\partial \bar{\sigma}}\right)^{\frac{d-2}{2}} \Theta[\bar{\sigma}]  .
\end{align}
For odd $d \geq 3$, it is instead
\begin{align}
\label{GreensFunction_Minkowski_Odd}
\bar{\mathcal{G}}_{\text{odd }d} \left[\sbar\right] 
= \frac{1}{\sqrt{2}(2\pi)^{\frac{d-1}{2}}} \left( \frac{ \partial }{ \partial \bar{\sigma} } \right)^{\frac{d-3}{2}} \left( \frac{\Theta[\bar{\sigma}]}{\sqrt{\bar{\sigma}}} \right) .
\end{align}
The embedding perspective we are exploring here has in fact been exploited in \cite{SoodakTiersten} to obtain the recursion relations in equations \eqref{GreensFunction_Minkowski_Even} and \eqref{GreensFunction_Minkowski_Odd}, relating $\bar{\mathcal{G}}_d$ and $\bar{\mathcal{G}}_{d+2}$. We will give a brief review of it here, by phrasing it as an application of the general formula eq. \eqref{GreensFunction_GeneralFormula}. 

Let the $(d+1)$D flat metric be $\eta_{\mu\nu}$ and the $(d+2)$D metric be $\eta_{\mathfrak{A}\mathfrak{B}}$, so that
\begin{align}
\eta_{\mathfrak{A}\mathfrak{B}} \dd X^\mathfrak{A} \dd X^\mathfrak{B}
= \eta_{\mu\nu} \dd X^\mu \dd X^\nu - \left(\dd X^{d+1}\right)^2 .
\end{align}
Comparison against eq. \eqref{Metric_GeneralEmbeddingForm} tells us $P = 1$ and $g^\perp_\text{AB} \dd y^\text{A} \dd y^\text{B} = -\left(\dd X^{d+1}\right)^2$. The zero mode equation in eq. \eqref{ZeroModeEquation} is
\begin{align}
\left(\frac{\dd}{\dd X^{d+1}}\right)^2 \psi_0 = 0,
\end{align}
whose general solution is
\begin{align}
\psi_0 = C_0 + C_1 X^{d+1} , \qquad C_{0,1} = \text{constant} .
\end{align}
We will choose the regular solution by setting $C_1 = 0$. There is no need to determine $C_0$ because it cancels out in the formula \eqref{GreensFunction_GeneralFormula}, which now hands us
\begin{align}
\label{GreensFunction_IntegralRecursionBetween_d_and_d+1}
\overline{G}^\pm_{d+1}[X-X']
&= \int_{-\infty}^{+\infty} \dd\left(X^{d+1}-X'^{d+1}\right)\overline{G}^\pm_{d+2}[X-X'] .
\end{align}
The same relation must hold if we now shift $d\to d-1$. This means
\begin{align}
\label{GreensFunction_IntegralRecursionBetween_d_and_d+2}
\overline{G}^\pm_d[X-X']
&= \int_{-\infty}^{+\infty} \dd\left(X^{d+1}-X'^{d+1}\right) \int_{-\infty}^{+\infty} \dd\left(X^d-X'^d\right) \overline{G}^\pm_{d+2}[X-X'] .
\end{align}
Next we re-write the Green's function so that it depends on only two variables, the time elapsed between observation and emission $T \equiv X^0-X'^0$ and the Euclidean distance between observer and source $R \equiv |\vec{X}-\vec{X}'|$. Using cylindrical symmetry on the $(X^{d},X^{d+1})$ plane and rotation symmetry of the $(X^1,\dots,X^{d-1})$ volume, we may convert eq. \eqref{GreensFunction_IntegralRecursionBetween_d_and_d+2} into
\begin{align}
\overline{G}^\pm_d[T,R]
&= 2\pi \int_R^\infty \dd R' R' \overline{G}^\pm_{d+2}[T,R'] .
\end{align}
Upon differentiating both sides with respect to $R$,
\begin{align}
\label{GreensFunction_DifferentialRecursionBetween_d_and_d+2}
-\frac{1}{2 \pi R} \frac{\partial}{\partial R } \overline{G}^\pm_d[T,R] &= \overline{G}^\pm_{d+2}[T,R] .
\end{align}
Global Poincar\'{e} symmetry informs us that, in fact, the Green's function depends solely on the object $\sbar$ in eq. \eqref{Minkowski_SyngeWorldFunction}; the $\Theta[\pm T]$ in \eqref{GreensFunction_Minkowski_RetardedAdvanced} is merely the instruction to ignore half of the light cone of $x'$. Because $(-(2\pi R)^{-1}\partial \sbar/\partial R) \partial/\partial \sbar = (2\pi)^{-1} \partial/\partial \sbar$, by induction on the number of dimensions $d$, the equivalence between eq. \eqref{GreensFunction_DifferentialRecursionBetween_d_and_d+2} and equations \eqref{GreensFunction_Minkowski_Even} and \eqref{GreensFunction_Minkowski_Odd} follows once the $d=2,3$ cases have been verified.

{\bf Euclidean} \qquad The flat Euclidean space Green's function for all spatial dimensions $d \geq 1$ reads
\begin{align}
\label{GreensFunction_Euclidean}
\overline{G}_d^\text{(E)}\left[\vec{X}-\vec{X}'\right] = -\frac{\Gamma\left[\frac{d}{2}-1\right]}{4\pi^{d/2} \left\vert\vec{X}-\vec{X}'\right\vert^{d-2}} .
\end{align}
($\Gamma$ is the Gamma function.) This formula is valid even for $d=2$, where the Green's function is proportional to $\ln| \vec{X} - \vec{X}' |$, if one first sets $d=2-\epsilon$ and proceed to expand in powers of $|\epsilon| \ll 1$ up to $\mathcal{O}[\epsilon^0]$. (Dimensional regularization acts as a long distance regulator here.) For $d=1$, $|X-X'|$ is to be read as the absolute value of the difference between the coordinates of the observer and source.

In this embedding framework, we see that all the Minkowski and Euclidean Green's functions follow from the $d=3$ flat spacetime case via integration and differentiation. Once a concrete expression is gotten for $\overline{G}^\pm_3[X-X']$, say by evaluating its Fourier integral representation, then $\overline{G}^\pm_2[X-X']$ can be obtained by integrating it once, using eq. \eqref{GreensFunction_IntegralRecursionBetween_d_and_d+1}. All even dimensional Green's functions then follow from $\overline{G}^\pm_2[X-X']$ by applying the differential recursion relation in eq. \eqref{GreensFunction_DifferentialRecursionBetween_d_and_d+2} repeatedly; and all odd dimensions from $\overline{G}^\pm_3[X-X']$. 

Since Euclidean space (with the Kronecker delta $\delta_{ij}$ as its metric) can be viewed as a $d$ dimensional space embedded in $(d+1)$ dimensional Minkowski, with the analog of eq. \eqref{Metric_GeneralEmbeddingForm} reading
\begin{align}
-\dd s^2 = \delta_{ij} \dd X^i \dd X^j - \left(\dd X^0\right)^2 , \qquad
P = 1, \qquad g^\perp_\text{AB} \dd y^\text{A} \dd y^\text{B} = - \left(\dd X^0\right)^2,
\end{align}
that means the Euclidean Green's function in eq. \eqref{GreensFunction_Euclidean} can be viewed as the massless minimally coupled scalar field generated by a line source in $(d+1)$D Minkowski spacetime, i.e., a static point source $\propto \delta^{(d)}[\vec{X}-\vec{X}']$, sweeping out a timelike worldline. Upon solving the zero mode equation in eq. \eqref{ZeroModeEquation}, which again yields a constant for the regular solution, formula \eqref{GreensFunction_GeneralFormula} now allows us to see that -- once the general $\overline{G}^\pm_{d+1}[X-X']$ is known, its Euclidean counterpart can be worked out with a single integration \cite{FourierTransform}
\begin{align}
\label{GreensFunction_EuclideanFromMinkowski}
\overline{G}_d^\text{(E)}\left[\vec{X}-\vec{X}'\right]
= \int_{-\infty}^\infty \dd\left(X^0-X'^0\right) \overline{G}^\pm_{d+1}[X-X'] .
\end{align}

\subsection{Lorenz gauge photons and de Donder gauge gravitons}

Before moving on to de Sitter spacetime, we remark that there is no difficulty in extending the embedding perspective for Minkowski massless scalar Green's functions to those of the Lorenz gauge photon and de Donder gauge graviton. This is because, in $(d>2)$-dimensional flat spacetime, these latter Green's functions are their massless scalar cousins multiplied by tensorial structures involving only $\eta_{\mu\nu}$ and parallel propagators $\eta_{\mu\nu'} = \eta_{\mu\nu}$. These objects, in turn, can be readily obtained from their $(d+1)$-dimensional counterparts by an appropriate projection.

We first record the Lorenz gauge ($\partial^\mu A_\mu = \partial^\mathfrak{B} A_\mathfrak{B} = 0$) photon Green's function in $d$- and $(d+1)$-Minkowski:
\begin{align}
\overline{G}^{(d)}_{\mu\nu'}[X-X'] &= \eta_{\mu\nu'} \overline{G}_d[X-X'], \\
\overline{G}^{(d+1)}_{\mathfrak{A}\mathfrak{B}'}[X-X'] &= \eta_{\mathfrak{A}\mathfrak{B}'} \overline{G}_{d+1}[X-X']  .
\end{align}
The de Donder gauge $(0 = \partial^\mu h_{\mu\nu} - (1/2) \partial_\nu h = \partial^\mathfrak{A} h_{\mathfrak{A}\mathfrak{B}} - (1/2) \partial_\mathfrak{B} h)$ graviton Green's function in $d$ and $(d+1)$-Minkowski are:
\begin{align}
\overline{G}^{(d)}_{\alpha\beta \mu'\nu'}[X-X'] 
	&= \frac{1}{2} \left( \eta_{\alpha\mu'} \eta_{\beta\nu'} + \eta_{\beta\mu'} \eta_{\alpha\nu'} 
	- \frac{2}{d-2} \eta_{\alpha\beta} \eta_{\mu'\nu'} \right) \overline{G}_d[X-X'] , \\
\overline{G}^{(d+1)}_{\mathfrak{A}\mathfrak{B} \mathfrak{M}'\mathfrak{N}'}[X-X'] 
	&= \frac{1}{2} \left( \eta_{\mathfrak{A}\mathfrak{M}'} \eta_{\mathfrak{B}\mathfrak{N}'} + \eta_{\mathfrak{B}\mathfrak{M}'} \eta_{\mathfrak{A}\mathfrak{N}'} 
	- \frac{2}{d-1} \eta_{\mathfrak{A}\mathfrak{B}} \eta_{\mathfrak{M}'\mathfrak{N}'} \right) \overline{G}_{d+1}[X-X'] .
\end{align}
If we define
\begin{align}
\varepsilon^\mathfrak{A}_{\phantom{\mathfrak{A}}\mu} &\equiv \frac{\partial X^\mathfrak{A}}{\partial X^\mu} ,
\end{align}
we see that
\begin{align}
\varepsilon^\mathfrak{A}_{\phantom{\mathfrak{A}}\mu} &= \delta^\mathfrak{A}_\mu \left(1 - \delta^\mathfrak{A}_d\right) .
\end{align}
Then, we may observe that, in $d$ dimensional Minkowski, both the parallel propagator and the Minkowski metric can be gotten from its $(d+1)$ dimensional cousin via
\begin{align*}
\eta_{\mu\nu'} = \eta_{\mathfrak{A}\mathfrak{B}'} 
\varepsilon^\mathfrak{A}_{\phantom{\mathfrak{A}}\mu} \varepsilon^{\mathfrak{B}'}_{\phantom{\mathfrak{B}'}\nu'} , \qquad
\eta_{\mu\nu} = \eta_{\mathfrak{A}\mathfrak{B}} 
\varepsilon^\mathfrak{A}_{\phantom{\mathfrak{A}}\mu} \varepsilon^{\mathfrak{B}}_{\phantom{\mathfrak{B}'}\nu} .
\end{align*}
We may immediately write down the analog of eq. \eqref{GreensFunction_IntegralRecursionBetween_d_and_d+1} for the Lorenz gauge photon Green's function,
\begin{align}
\label{Embedding_LorenzGaugePhoton}
\overline{G}^{(d)}_{\mu\nu'}[X-X'] 
= \int_{-\infty}^{+\infty} \dd X'^d \overline{G}^{(d+1)}_{\mathfrak{A}\mathfrak{B}'}[X-X'] \varepsilon^\mathfrak{A}_{\phantom{\mathfrak{A}}\mu} \varepsilon^{\mathfrak{B}'}_{\phantom{\mathfrak{B}'}\nu'} ;
\end{align}
whereas that for the de Donder gauge  graviton Green's function reads
\begin{align}
\label{Embedding_deDonderGaugeGraviton}
\overline{G}^{(d)}_{\alpha\beta \mu'\nu'}[X-X'] 
= \int_{-\infty}^{+\infty} \dd X'^d \overline{G}^{(d+1)}_{\mathfrak{A}\mathfrak{B} \mathfrak{M}'\mathfrak{N}'}[X-X'] \varepsilon^\mathfrak{A}_{\phantom{\mathfrak{A}}\alpha} \varepsilon^\mathfrak{B}_{\phantom{\mathfrak{B}}\beta} 
\varepsilon^{\mathfrak{M}'}_{\phantom{\mathfrak{M}'}\mu'} \varepsilon^{\mathfrak{N}'}_{\phantom{\mathfrak{N}'}\nu'} .
\end{align}

\section{de Sitter Spacetime}
\label{Section_deSitter}

de Sitter spacetime in $d$ dimensions is defined as the following hyperboloid embedded in $(d+1)$ dimensional Minkowski:
\begin{align}
-\eta_{\mathfrak{A}\mathfrak{B}} X^\mathfrak{A} X^\mathfrak{B} \equiv -X^2 = \frac{1}{H^2} , \qquad H > 0 .
\end{align}
$H$ is some fixed Hubble parameter. Let $\tau\in\mathbb{R}$, $\rho \geq 0$, $\{\theta^i\}$ be the $(d-1)$ angular coordinates on a $(d-1)$-sphere, and $\widehat{n}[\vec{\theta}]$ be the unit radial (spatial) vector. For the most part we will employ hyperbolic coordinates to describe an arbitrary point in Minkowski spacetime outside the light cone of its origin $0^\mathfrak{A}$,
\begin{align}
\label{deSitter_X_parametrization}
X^\mathfrak{A}[\rho,\tau,\vec{\theta}] = \rho \left( \sinh[\tau], \cosh[\tau] \widehat{n}[\vec{\theta}] \right) .
\end{align}
The Minkowski metric, for $X^2 < 0$, now reads
\begin{align}
\label{Embedding_deSitter}
\dd s^2 &= \rho^2 \left( \dd\tau^2 - \cosh^2[\tau] \dd\Omega_{d-1}^2 \right) - \dd\rho^2 ,
\end{align}
with $\dd\Omega_{d-1}^2$ being the metric on the $(d-1)$D sphere -- such that the induced metric on the $\rho=1/H$ surface is de Sitter spacetime
\begin{align}
g_{\mu\nu}^\text{(dS)} \dd x^\mu \dd x^\nu
= \frac{1}{H^2} \left( \dd\tau^2 - \cosh^2[\tau] \dd\Omega_{d-1}^2 \right) .
\end{align}
These coordinates, which cover the whole of de Sitter spacetime, are also known as the ``closed slicing" coordinates because constant time surfaces describe a closed sphere: $(g_{\mu\nu}^\text{(dS)} \dd x^\mu \dd x^\nu)_{\tau \text{ fixed}} \propto \dd\Omega_{d-1}^2$.

Our main results will be expressed in terms of the object
\begin{align}
\label{deSitter_Z_Definitions}
Z[x,x'] \equiv \left.\Big( H^2 X\left[ \rho, x \right] \cdot X'\left[ \rho', x' \right] \Big)\right\vert_{\rho=\rho'=H^{-1}}.
\end{align}
With the choice of coordinates in eq. \eqref{deSitter_X_parametrization}, for instance, we have
\begin{align}
Z\left[ \tau,\vec{\theta};\tau',\vec{\theta}' \right] = \sinh[\tau] \sinh[\tau'] - \cosh[\tau] \cosh[\tau'] \widehat{n}\cdot\widehat{n}' .
\end{align}
We also note that the square of the geodesic distance between $x$ and $x'$ in de Sitter is 
\begin{align}
L[x,x'] = \left(\frac{1}{H} \cosh^{-1}\Big[-Z[x,x']\Big] \right)^2.
\end{align}
In particular, $x$ and $x'$ lie precisely on or within each other's null cones, $L[x,x'] \geq 0$, and are thus causally connected, when and only when
\begin{align}
\label{deSitter_CausalCondition}
Z[x,x'] \leq -1 .
\end{align}

{\bf Integral representation} \qquad With these preliminaries aside, we may now apply the general formula \eqref{GreensFunction_GeneralFormula} to eq. \eqref{Embedding_deSitter}. First we identify
\begin{align}
P[\rho] = \rho, \qquad g_\text{AB}^\perp \dd y^\text{A} \dd y^\text{B} = -\dd\rho^2 .
\end{align}
The zero mode equation is
\begin{align}
\D_\rho \psi_0 = -\frac{1}{\rho^{d-2}} \frac{\dd}{\dd\rho} \left( \rho^d \frac{\dd \psi_0}{\dd\rho} \right) = 0,
\end{align}
and its general solution is
\begin{align}
\psi_0[\rho] = C_1 \rho^{1-d} + C_2 , \qquad C_{1,2} = \text{constant.}
\end{align}
We will choose the regular solution by putting $C_1 = 0$. There is no need to determine $C_2$ since it cancels out in eq. \eqref{GreensFunction_GeneralFormula}.

The integral representation of the minimally coupled massless scalar Green's function in de Sitter spacetime is therefore
\begin{align}
\label{GreensFunction_deSitter_IntegralRepresentation}
G_d[x,x']
= \int_0^\infty \dd\rho' \left(H \rho'\right)^{d-2} \overline{G}_{d+1} \Big[ X[\rho=H^{-1},x] - X'[\rho',x'] \Big] ,
\end{align}
with the parametrization in eq. \eqref{deSitter_X_parametrization}, and the flat spacetime Green's functions $\overline{G}_{d+1}$ from equations \eqref{GreensFunction_Minkowski_RetardedAdvanced} through \eqref{GreensFunction_Minkowski_Odd}. As already advertised, this admits the interpretation that the de Sitter scalar Green's function is sourced by a line charge which begins from one end infinitesimally close to the Minkowski origin $0^\mathfrak{A}$, penetrates perpendicularly the de Sitter hyperboloid at $X'[\rho'=1/H,x']$, and continues to infinity. (One can check directly, from eq. \eqref{deSitter_X_parametrization}, that the vector $\partial_{\rho'} X'^{\mathfrak{A}}$ is orthogonal to $\partial_{\tau'} X'^{\mathfrak{A}}$ and $\{\partial_{\theta'^i} X'^{\mathfrak{A}}\}$.) The observer constrained to live on the $d$ dimensional hyperboloid at $\rho=1/H$ has been deceived to think she sees a $\delta$-function point source at $(\rho' = 1/H, x')$.

{\bf Causal structure for closed slicing} \qquad Before we compute eq. \eqref{GreensFunction_deSitter_IntegralRepresentation}, we need to first tackle an issue we have already raised in section \eqref{Section_Generalities}. If a family of observers living in de Sitter spacetime held clocks synchronized to read $\tau$ occurring in the parametrization of eq. \eqref{deSitter_X_parametrization}, and if cause preceded effect (or effect preceded cause) from the higher dimensional Minkowski viewpoint, then does cause necessarily precede effect (effect precede cause) from these de Sitter observers' viewpoint? In Minkowski spacetime itself, for instance, whether cause or effect comes first is controlled by the $\Theta[\pm(X^0-X'^0)]$ in eq. \eqref{GreensFunction_Minkowski_RetardedAdvanced}. We now answer the above question in the affirmative, and demonstrate that, the retarded $G_d^+$ (advanced $G_d^-$) Green's function in the closed slicing coordinates of de Sitter, is obtained simply from its retarded (advanced) counterpart in Minkowski:
\begin{align}
\label{GreensFunction_deSitter_CausalityOverlapsWithMinkowskiInClosedSlicing}
G_d^{(\text{Closed}\vert\pm)}[x,x'] &= \int_0^\infty \dd\rho' \left(H \rho'\right)^{d-2} 
		\Theta\left[ \pm(X^0-X'^0) \right] \bar{\mathcal{G}}_{d+1} \Big[ X[\rho,\tau,\widehat{n}] - X'[\rho',\tau',\widehat{n}'] \Big] \\
&= \Theta[\pm(\tau-\tau')] H^{d-2} \mathcal{G}_d[x,x'] ,
\end{align}
where $\bar{\mathcal{G}}_{d+1}$ and $\mathcal{G}_d$ are both symmetric (unordered in time); the former can be found in equations \eqref{GreensFunction_Minkowski_Even} and \eqref{GreensFunction_Minkowski_Odd}, and the latter is the analog of eq. \eqref{GreensFunction_GeneralFormula_Check_Integral}:
\begin{align}
\label{GreensFunction_deSitter_SymmetricDefinition}
\mathcal{G}_d[x,x'] &\equiv \int_0^\infty \dd\rho' \rho'^{d-2} \bar{\mathcal{G}}_{d+1} \Big[ X[\rho,\tau,\widehat{n}] - X'[\rho',\tau',\widehat{n}'] \Big] .
\end{align}
(We have left $\rho$ unevaluated so that we can exercise the check discussed around equations \eqref{GreensFunction_GeneralFormula_Check_IofII} through \eqref{GreensFunction_GeneralFormula_Check_Integral}.) The key point here is that, for a given observer location $(\rho=1/H,\tau,\widehat{n})$ and some de Sitter point source \cite{SourcesConfusion} position $(\rho'=1/H,\tau',\widehat{n}')$,
\begin{align}
\label{deSitter_Causal_StepFunctions}
\Theta\left[\pm\left( X^0[\rho,\tau,\widehat{n}] - X'^0[\rho',\tau',\widehat{n}'] \right)\right] =
\Theta\left[ \pm(\tau-\tau') \right]
\end{align}
for all $\rho'$ such that $(X-X')^2 \geq 0$. This follows from the fact that the sign of the time component of a timelike or null vector in Minkowski spacetime is a Lorentz invariant; namely, $\tau-\tau' \equiv V^0$ for $V \equiv X/\sqrt{-X^2} - X'/\sqrt{-X'^2}$, with $V^2 \geq 0$ whenever $(X-X')^2 \geq 0$.

\begin{figure}
\begin{center}
\includegraphics[width=4in]{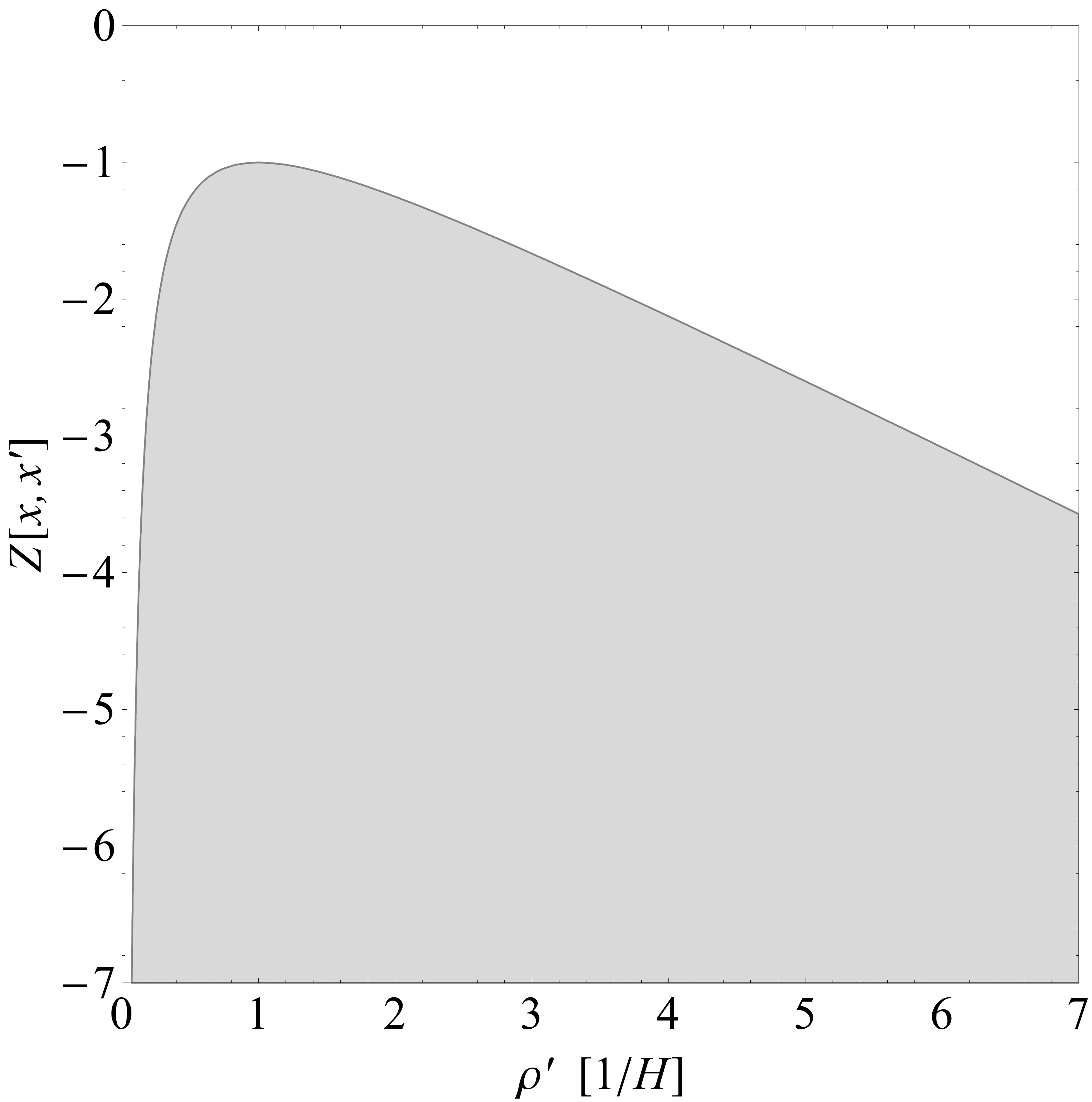}
\caption{The light gray region is the range of $Z[x,x']$ and $\rho'$ contributing to the signal at the observer's location on the de Sitter hyperboloid (i.e., at $X[\rho=H^{-1},x]$), as described by the symmetric Green's function of eq. \eqref{GreensFunction_deSitter_SymmetricDefinition}. The dimensionless $Z[x,x']$ is defined in eq. \eqref{deSitter_Z_Definitions}. The point $(\rho',Z) = (1/H,-1)$ is responsible for the signal propagating on the light cone. Whereas for a fixed $Z < -1$, the range of $\rho'$ lying within the light gray region quantifies which segment of the line source, parametrized in eq. \eqref{deSitter_X_parametrization}, produced the scalar wave tails seen by the observer at $x$. Here, we are expressing $\rho'$ in units of $1/H$, the size of the de Sitter hyperboloid.}
\label{Figure_CausallyConnected}
\end{center}
\end{figure}

That the observer at $X$ and the source at $X'$ has to be within each other's light cone (in the Minkowski sense), is imposed directly by the flat Green's functions in eq. \eqref{GreensFunction_deSitter_CausalityOverlapsWithMinkowskiInClosedSlicing}. (See the occurrence of $\Theta[\pm(X^0-X'^0)] \Theta[\sbar]$ in equations \eqref{GreensFunction_Minkowski_RetardedAdvanced} through \eqref{GreensFunction_Minkowski_Odd}.) With the parametrization in eq. \eqref{deSitter_X_parametrization}, this translates to
\begin{align}
\label{deSitter_CausalCondition_MinkowskiSense}
(X-X')^2 = -\left( \rho^2 + \rho'^2 + 2 \rho\rho' Z \right) \geq 0 ,
\end{align}
which, in turn, holds when and only when \cite{CausalStructureAlgebra}
\begin{align}
\label{deSitter_CausalCondition_MinkowskiSense_rho'Limits}
-Z - \sqrt{Z^2-1} \leq \frac{\rho'}{\rho} \leq -Z + \sqrt{Z^2-1} , \qquad \text{and} \qquad Z \leq -1 .
\end{align}
Notice we have recovered the causality condition $Z[x,x'] \leq -1$ in eq. \eqref{deSitter_CausalCondition}; it is apparently encoded in the integral representation eq. \eqref{GreensFunction_deSitter_IntegralRepresentation}. From the Minkowski point of view, the light cone part of the signal detected by the observer originates from the point on the line source intersecting with the de Sitter hyperboloid (at $\rho'/\rho=-Z=1$); while the tail part of the signal comes from the rest of the line. Fig. \eqref{Figure_CausallyConnected} illustrates the conditions imposed by eq. \eqref{deSitter_CausalCondition_MinkowskiSense_rho'Limits}.

We have framed our derivation of the retarded and advanced de Sitter Green's functions in terms of the closed slicing coordinates in eq. \eqref{deSitter_X_parametrization}. It is important to note that, however, because the object $\Theta[\pm(\tau-\tau')]$ is written in terms of the time coordinates in a particular ``slicing" of de Sitter, it is not a coordinate scalar. The practical strategy we will adopt here is to focus on the symmetric Green's function in eq. \eqref{GreensFunction_deSitter_SymmetricDefinition}. As we will shortly demonstrate, it depends solely on $Z$ and therefore can be viewed in a manner independent of the parametrization chosen on de Sitter itself. Only when one does pick a particular set of coordinates -- for example, the flat slicing in eq. \eqref{deSitter_4D_FlatSlicing} below -- do we then multiply it by $\Theta[\pm(t-t')]$ (where $t$ and $t'$ are the time coordinates), to get the retarded $(+)$ and advanced $(-)$ Green's functions.

{\bf Technicalities} \qquad We now turn to performing the integral in eq. \eqref{GreensFunction_deSitter_SymmetricDefinition}. First rescale the integration variable $\rho'' \equiv \rho'/\rho$, which effectively sets $\rho=1$, so that now $\sbar = -(1/2)(\rho''^2 + 1 + 2 \rho'' Z)$. Because the flat spacetime symmetric Green's functions depend only on $\sbar$, we may proceed to replace in equations \eqref{GreensFunction_Minkowski_Even} and \eqref{GreensFunction_Minkowski_Odd}
\begin{align}
\frac{\partial}{\partial \sbar} \to -\frac{1}{\rho''} \frac{\partial}{\partial Z} .
\end{align}
The limits of $\rho''$ integration are determined by the causal condition $(X-X')^2 \geq 0$ (from the $\Theta[\sbar]$ in in equations \eqref{GreensFunction_Minkowski_Even} and \eqref{GreensFunction_Minkowski_Odd}); the result has already been found in eq. \eqref{deSitter_CausalCondition_MinkowskiSense_rho'Limits}. For $Z > -1$ the integral is to be set to zero. At this point, the even and odd $d$ integrals are respectively
\begin{align}
\label{GreensFunction_deSitter_Even_StepI}
\mathcal{G}_{\text{even }d}[x,x']
&= \frac{1}{(2\pi)^{\frac{d}{2}}} \left( -\frac{\partial}{\partial Z} \right)^{\frac{d-2}{2}} \left( \Theta[-Z-1] 
\int_{-Z-\sqrt{Z^2-1}}^{-Z+\sqrt{Z^2-1}} \frac{\dd \rho'' \rho''^{\frac{d-2}{2}}}{\sqrt{-\left(\rho''^2+1+2\rho''Z\right)}} 
\right)
\end{align}
and
\begin{align}
\label{GreensFunction_deSitter_Odd_StepI}
\mathcal{G}_{\text{odd }d}[x,x']
&= \frac{1}{2(2\pi)^{\frac{d-1}{2}}} \left( -\frac{\partial}{\partial Z} \right)^{\frac{d-1}{2}} 
		\left( \Theta[-Z-1] \int_{-Z-\sqrt{Z^2-1}}^{-Z+\sqrt{Z^2-1}} \dd\rho'' \rho''^{\frac{d-1}{2}-1} \right) .
\end{align}
Notice equations \eqref{GreensFunction_deSitter_Even_StepI} and \eqref{GreensFunction_deSitter_Odd_StepI} have both become independent of $\rho$, the size of the observer's de Sitter hyperboloid. Referring back to the discussion around eq. \eqref{GreensFunction_GeneralFormula_Check_Integral}, we may conclude these are indeed the (symmetric) de Sitter Green's functions.

Since the integral in eq. \eqref{GreensFunction_deSitter_Odd_StepI} for odd $d$ involves merely a power of $\rho''$, the remaining challenge is that of the even $d$ case in eq. \eqref{GreensFunction_deSitter_Even_StepI}. We massage it by changing variables to $\rho' \equiv -Z + \cos[\varphi] \sqrt{Z^2-1}$, thereby bringing us to
\begin{align}
\int_{-Z-\sqrt{Z^2-1}}^{-Z+\sqrt{Z^2-1}} \frac{\dd \rho'' \rho''^{\frac{d-2}{2}}}{\sqrt{-\left(\rho''^2+1+2\rho''Z\right)}} 
= \int_0^\pi \dd\varphi \left((-Z) + \sqrt{(-Z)^2-1} \cos\varphi\right)^{\frac{d-2}{2}} .
\end{align}
A glance at eq. (8.913.3) of \cite{G&S} tells us this is $\pi$ times the Legendre polynomial of degree $(d-2)/2$, $P_{\frac{d-2}{2}}[-Z]$.

{\bf Results} \qquad To sum, the retarded $G_d^+$ and advanced $G_d^-$ Green's functions of the minimally coupled massless scalar field, in the closed slicing coordinates of eq. \eqref{deSitter_X_parametrization}, are given by $G_d^{(\text{Closed}\vert\pm)}[x,x'] = \Theta[\pm(\tau-\tau')] H^{d-2} \mathcal{G}_d[x,x']$. The symmetric Green's functions $\mathcal{G}_d$, in turn, are expressed in terms of $Z[x,x']$ defined in eq. \eqref{deSitter_Z_Definitions},
\begin{align}
\label{GreensFunction_deSitter_Even}
\mathcal{G}_{\text{even }d}[x,x']
&= \frac{\pi}{(2\pi)^{\frac{d}{2}}} \left( -\frac{\partial}{\partial Z} \right)^{\frac{d-2}{2}} \left( \Theta[-Z-1] P_{\frac{d-2}{2}}\left[-Z\right]  \right), \\
\label{GreensFunction_deSitter_Odd}
\mathcal{G}_{\text{odd }d}[x,x']
&= \frac{1}{(d-1)(2\pi)^{\frac{d-1}{2}}} \left( -\frac{\partial}{\partial Z} \right)^{\frac{d-1}{2}} \\
&\qquad \times
\left( \Theta[-Z-1] \left\{ \left(-Z+\sqrt{Z^2-1}\right)^{\frac{d-1}{2}}-\left(-Z-\sqrt{Z^2-1}\right)^{\frac{d-1}{2}} \right\} \right) . \nonumber
\end{align}
The $P_{\frac{d-2}{2}}[-Z]$ is the Legendre polynomial and the terms inside the curly brackets in eq. \eqref{GreensFunction_deSitter_Odd} can be re-expressed in terms of $\sqrt{Z^2-1} U_{\frac{d-1}{2}}[-Z]$, with $U_{\frac{d-1}{2}}[-Z]$ being Chebyshev's polynomial of the second kind -- see eq. (8.940) of \cite{G&S}.

The presence of $\Theta[-Z-1]$ in equations \eqref{GreensFunction_deSitter_Even} and \eqref{GreensFunction_deSitter_Odd} is necessary to impose the $Z[x,x'] \leq -1$ causality condition, that $x$ and $x'$ lie within each other's light cone (in the de Sitter sense). Furthermore -- again from the de Sitter perspective -- terms that contain $(-\partial_Z) \Theta[-Z-1] = \delta[Z+1]$ and higher derivatives of the $\delta$-function describe propagation of signals on the light cone, whereas the terms containing $\Theta[-Z-1]$, namely
\begin{align}
\label{GreensFunction_deSitter_Even_Tail}
\mathcal{G}_{\text{even }d}^{\text{(Tail)}}[x,x']
&= \frac{\pi \Theta[-Z-1]}{(2\pi)^{\frac{d}{2}}} \frac{(d-2)!}{2^{\frac{d-2}{2}} \left(\frac{d-2}{2}\right)!}, \\
\label{GreensFunction_deSitter_Odd_Tail}
\mathcal{G}_{\text{odd }d}^{\text{(Tail)}}[x,x']
&= \frac{\Theta[-Z-1]}{(d-1)(2\pi)^{\frac{d-1}{2}}} 
\left( -\frac{\partial}{\partial Z} \right)^{\frac{d-1}{2}} \left\{ \left(-Z+\sqrt{Z^2-1}\right)^{\frac{d-1}{2}}-\left(-Z-\sqrt{Z^2-1}\right)^{\frac{d-1}{2}} \right\} , 
\end{align}
describe the tail effect. Despite its zero mass, a portion of the scalar field propagates inside the light cone of the source \cite{RodriguesTail}.

{\bf Four dimensions} \qquad There is mounting observational evidence that our universe underwent a period of accelerated expansion during the very earliest moments of its existence, and is currently entering a similar ``dark energy" dominated phase. To zeroth order, the metric of such a spacetime is de Sitter. The same observational evidence indicates that, what is most relevant, however, is not the closed slicing parametrization we chose in eq. \eqref{deSitter_X_parametrization}, where constant time surfaces of de Sitter are closed spheres -- but rather the flat slicing parametrization, where constant time surfaces are spatially flat and have infinite volume. This choice of coordinates is given by the formulas
{\allowdisplaybreaks\begin{align}
\label{deSitter_4D_FlatSlicing}
X^0 &= \frac{1}{2\eta} \left(\eta^2 - \vec{x}^2 - \frac{1}{H^2} \right) , \\
X^4 &= \frac{1}{2\eta} \left(-\eta^2 + \vec{x}^2 - \frac{1}{H^2} \right) , \nonumber\\
X^1 &= \frac{x^1}{H\eta}, \qquad X^2 = \frac{x^2}{H\eta}, \qquad X^3 = \frac{x^3}{H\eta} , \nonumber
\end{align}}
which in turn give us the following form of the de Sitter metric:
\begin{align}
\label{Metric_deSitter_4D_SpatiallyFlat}
g_{\mu\nu}^\text{(dS)} \dd x^\mu \dd x^\nu = \frac{\dd \eta^2 - \dd\vec{x}^2}{(H\eta)^2} .
\end{align}
Here, $-\infty < \eta < 0$ and $\vec{x}^2 \equiv \sum_{i=1}^{3} (x^i)^2$. In these coordinates,
\begin{align}
\label{Metric_deSitter_4D_-Z-1}
-Z-1 = \frac{1}{2\eta\eta'} \left( \left( \eta-\eta' \right)^2 - \left( \vec{x}-\vec{x}' \right)^2 \right) .
\end{align}
By setting $-Z\delta[Z+1] = \delta[Z+1]$, the symmetric Green's function in eq. \eqref{GreensFunction_deSitter_Even} takes the form
\begin{align}
\label{GreensFunction_deSitter_4D_IofIII}
\mathcal{G}_4[x,x']
&= \frac{1}{4\pi} \left( \delta[Z+1] + \Theta[-Z-1] \right), \qquad \text{(Coordinate independent)} \\
\label{GreensFunction_deSitter_4D_IIofIII}
&= \frac{1}{4\pi} \left( \eta\eta' \delta\left[ \mathcal{S} \right] + \Theta\left[ \mathcal{S} \right] \right), \qquad
\mathcal{S} \equiv \frac{1}{2}\left( \left( \eta-\eta' \right)^2 - \left( \vec{x}-\vec{x}' \right)^2 \right), \qquad \text{(Flat slicing)} .
\end{align}
To obtain the retarded and advanced Green's functions, we multiply by $\Theta[\pm(\eta-\eta')]$.
\begin{align}
\label{GreensFunction_deSitter_4D_IIIofIII}
G_4^{(\text{Flat}|\pm)}[x,x'] = \Theta[\pm(\eta-\eta')] H^2 \mathcal{G}_4[x,x'] .
\end{align}
These results are consistent with, for instance, the discussion in \cite{Burko:2002ge}.

\section{Topological Obstructions And Green's Functions On $d$-Sphere}
\label{Section_nSphere}

Having derived the de Sitter Green's functions from the Minkowski ones, let us now ask if there are potential obstructions to our general construction. In this section, we shall restrict our attention to Riemannian spaces with positive definite metrics. Recall that the Gauss-Stokes' theorem tells us the integral of the divergence of the gradient of a scalar field $\varphi$ over some volume, is equal to the flux of this gradient field over the boundary of the same volume. In a closed volume such the $d$-dimensional sphere, i.e., where there is no boundary, the integral of such a total divergence has to vanish identically. In physical terms, topology forbids a net positive charge on a $d$-sphere without a compensating negative charge: the total integral of the charge density $J \equiv \Box \varphi$ has to be zero.

Now, a $d$-sphere can be defined by the following embedding in $(d+1)$D Euclidean space:
\begin{align}
\sum_{\mathfrak{A}=1}^{d+1} \left(X^\mathfrak{A}\right)^2 \equiv \vec{X}^2 = R^2 , \qquad R > 0.
\end{align}
The $(d+1)$D Euclidean metric in spherical coordinates is
\begin{align}
\dd \vec{s}^2 = r^2 \dd\Omega_d^2 + \dd r^2 ,
\end{align}
where $\dd\Omega_d^2$ is the metric on the $d$-sphere. This is of the form in eq. \eqref{Metric_GeneralEmbeddingForm}, with the identifications
\begin{align}
P[r] = r, \qquad g^\perp_\text{AB} \dd y^\text{A} \dd y^\text{B} = \dd r^2.
\end{align}
An arbitrary point in the $(d+1)$D Euclidean space has coordinates
\begin{align}
\label{dSphere_X_parametrization}
\vec{X}[r,\widehat{n}] = r \widehat{n}[\vec{\theta}],
\end{align}
where $\widehat{n}[\vec{\theta}]$ is the unit radial vector parametrized by $d$ angular coordinates $\vec{\theta}$.

Let us press on with the application of the general formula in eq. \eqref{GreensFunction_GeneralFormula}, to see how it will break down. The zero mode equation is
\begin{align}
D_r \psi_0 = \frac{1}{r^{d-2}} \frac{\dd}{\dd r} \left( r^d \frac{\dd \psi_0}{\dd r} \right) = 0,
\end{align}
whose general solution is
\begin{align}
\psi_0[r] = C_1 r^{1-d} + C_2 .
\end{align}
Choosing the regular solution, we set $C_1=0$. The general formula in eq. \eqref{GreensFunction_GeneralFormula} becomes 
\begin{align}
\label{dSphere_Step0}
G_d^{(\text{E}|+)}[\widehat{n},\widehat{n}';r] 
&= \int_0^\infty \dd r' \left(\frac{r'}{R}\right)^{d-2} \overline{G}_{d+1}^\text{(E)}\Big[ \vec{X}[r,\widehat{n}] - \vec{X}'[r',\widehat{n}'] \Big] \\
&\equiv -\frac{\Gamma\left[\frac{d-1}{2}\right]}{4\pi^{\frac{d+1}{2}} R^{d-2}} \mathcal{G}_d^{(\text{E}|+)}[\widehat{n},\widehat{n}';r] 
\end{align}
where $\overline{G}_{d+1}^\text{(E)}$ can be found in eq. \eqref{GreensFunction_Euclidean}; and we have extracted the analog of eq. \eqref{GreensFunction_GeneralFormula_Check_Integral} for evaluation
\begin{align}
\label{dSphere_StepI}
\mathcal{G}_d^{(\text{E}|+)}[\widehat{n},\widehat{n}';r] 
&\equiv \int_0^\infty \frac{\dd r' r'^{d-2}}{\left\vert r^2 + r'^2 - 2 r r' \widehat{n}\cdot\widehat{n}' \right\vert^{\frac{d-1}{2}}} .
\end{align}
We will eventually set $r=R$, but just as in the previous section, we will leave it unevaluated for now. At first sight, one may be tempted to re-scale $r'' \equiv r'/r$ in eq. \eqref{dSphere_StepI}, and conclude that the resulting integral is independent of $r$. However, the situation is more subtle because eq. \eqref{dSphere_StepI} is ``logarithmically divergent":
\begin{align}
\label{dSphere_StepII}
\lim_{r_\text{IR} \to \infty} \int^{r_\text{IR}} \frac{\dd r' r'^{d-2}}{\left\vert r^2 + r'^2 - 2 r r' \widehat{n}\cdot\widehat{n}' \right\vert^{\frac{d-1}{2}}} 
= \int^{r_\text{IR}} \frac{\dd r' r'^{d-2}}{r'^{d-1}} 
= \ln[r_\text{IR}] .
\end{align}
From the following indefinite integrals, with integers $m=0,1,2,3,\dots$,
\begin{align}
\int z^{2m} (a z^2 + b z + c)^{-(1+2m)/2}\dx{z}
		&= \frac{(-)^m}{(1/2)_m}\partial_a^m \left( a^{-1/2} \ln\left[ 2 \sqrt{a} \sqrt{a z^2 + b z + c} + 2az + b \right] \right) \\
\int z^{1+2m} (a z^2 + b z + c)^{-1-m}\dx{z}
		&= \frac{(-)^m}{(1)_m}\partial_a^m \left( \frac{1}{2a} \left( 
	\ln\left[a z^2 + b z + c\right] 
	- \frac{2b}{\sqrt{4ac-b^2}} \tan^{-1}\left[\frac{2az+b}{\sqrt{4ac-b^2}}\right] \right) \right) ,
\end{align}
where $(\beta)_m \equiv \beta (\beta + 1) (\beta + 2) \dots (\beta + (m-1))$ is the Pochhammer symbol -- we may in fact understand \eqref{dSphere_StepI} to mean, for even $d \geq 2$ \cite{PochhammerCalculation},
\begin{align}
\label{dSphere_StepIII}
\lim_{r_\text{IR} \to \infty}\int_0^{r_\text{IR}} \dd r' & \frac{r'^{d-2}}{\left(r'^2 + r^2 - 2 r r' \widehat{n}\cdot\widehat{n}'\right)^{\frac{d-1}{2}}} \\
&= - \ln[r/r_\text{IR}] + \ln[2] 
	+ \left. \frac{(-)^{\frac{d-2}{2}}}{(1/2)_{\frac{d-2}{2}}} \partial_a^{\frac{d-2}{2}} \left( \frac{\ln\left[a\right] - \ln\left[\sqrt{a} - \widehat{n} \cdot \widehat{n}' \right]}{\sqrt{a}} \right) \right\vert_{a=1}  \nonumber
\end{align}
and, for odd $d \geq 3$,
\begin{align}
\label{dSphere_StepIV}
&\lim_{r_\text{IR} \to \infty}\int_0^{r_\text{IR}} \dd r' \frac{r'^{d-2}}{\left(r'^2 + r^2 - 2 r r' \widehat{n}\cdot\widehat{n}'\right)^{\frac{d-1}{2}}} \\
&= - \ln[r/r_\text{IR}] 
	+ \left. \frac{(-)^{\frac{d-3}{2}}}{(1)_{\frac{d-3}{2}}}\partial_a^{\frac{d-3}{2}} \left( \frac{1}{a} \left( \frac{\ln\left[a\right]}{2} 
	+ \frac{\widehat{n}\cdot\widehat{n}'}{\sqrt{a - \left(\widehat{n} \cdot \widehat{n}'\right)^2}} \left(\frac{\pi}{2} + \tan^{-1}\left[\frac{\widehat{n} \cdot \widehat{n}'}{\sqrt{a - \left(\widehat{n} \cdot \widehat{n}'\right)^2}}\right] \right) \right) \right) \right\vert_{a=1} . \nonumber
\end{align}
Notice the lower and upper (divergent) limits of eq. \eqref{dSphere_StepI} have conspired to yield a dependence on $r$ through $\ln[r/r_\text{IR}]$; if the integral had converged, as was the case for the de Sitter calculation, $r$ could have been re-scaled away. The situation is akin to that in quantum field theory, where the need to tame otherwise divergent calculations forces one to introduce extra dimension-ful scales in the problem.

Recalling the discussion around  eq. \eqref{GreensFunction_GeneralFormula_Check_Integral}, our integral representation in eq. \eqref{dSphere_StepI} is therefore not a valid solution to the Green's function of the Laplacian on the $d$-sphere, because $\D_r \ln[r/r_\text{IR}] \neq 0$. Of course no such solution should exist, for the topological reasons we have already mentioned. But the constraint that equal amounts of positive and negative charge need to exist on the $d$-sphere suggests that we can construct a modified Green's function sourced by one positive point charge located at $\widehat{n}_+$ on the sphere and one negative point charge at $\widehat{n}_-$, i.e.,
\begin{align}
\label{dSphere_ModifiedGreensFunction_Equation}
R^{d-2} \overline{\Box}_{\widehat{n}} G_d^\text{(E$\vert\pm$)}\left[ \widehat{n}\cdot\widehat{n}_+, \widehat{n}\cdot\widehat{n}_-; R\right] 
= \delta^{(d)}[\widehat{n}-\widehat{n}_+] - \delta^{(d)}[\widehat{n}-\widehat{n}_-],
\end{align}
where $\delta^{(d)}[\widehat{n}-\widehat{n}']$ is shorthand for the appropriate $\delta$-function measure. For instance, on the 2-sphere, we have $\delta^{(2)}[\widehat{n}-\widehat{n}'] = \delta[\cos\theta-\cos\theta'] \delta[\phi-\phi']$.

This modified solution is nothing but the difference between two of the ``Green's functions" in eq. \eqref{dSphere_Step0}, with one $\widehat{n}' \to \widehat{n}_+$ and the other $\widehat{n}' \to \widehat{n}_-$. That this difference has to solve eq. \eqref{dSphere_ModifiedGreensFunction_Equation}, is because all the pieces independent of $\widehat{n}$ and $\widehat{n}'$ in equations \eqref{dSphere_StepIII} and \eqref{dSphere_StepIV} cancel out. In particular, the cancellation of the $\ln[r/r_\text{IR}]$ simultaneously cures the logarithmic divergence and yields a $r$-independent answer: if one now applies $\overline{\Box}$ to this difference, the analog of eq. \eqref{GreensFunction_GeneralFormula_Check_IIofII} is
\begin{align}
R^{d-2} \overline{\Box}_{\widehat{n}} G_d^{(\text{E}|\pm)}[\widehat{n}\cdot\widehat{n}_+, \widehat{n}\cdot\widehat{n}_- ; R]
&= \delta^{(d)}[\widehat{n}-\widehat{n}_+] - \delta^{(d)}[\widehat{n}-\widehat{n}_-] \\
&\qquad\qquad
+ \left.\frac{\Gamma\left[\frac{d-1}{2}\right]}{4\pi^{\frac{d+1}{2}}} 
	\D_r\left( \mathcal{G}_d^{(\text{E}|+)}[\widehat{n},\widehat{n}_+;r] - \mathcal{G}_d^{(\text{E}|+)}[\widehat{n},\widehat{n}_-;r] \right)\right\vert_{r=R} ,
\nonumber
\end{align} 
i.e., the second line is zero.

{\bf Results} \qquad We have thus proposed the following modified Green's function of the Laplacian on the $d$-sphere of radius $R$, embedded in $(d+1)$D Euclidean space, obeying eq. \eqref{dSphere_ModifiedGreensFunction_Equation}. Its integral representation is
\begin{align}
\label{dSphere_ModifiedGreensFunction_IntegralRepresentation}
G_d^{(\text{E}|\pm)}[\widehat{n}\cdot\widehat{n}_+, \widehat{n}\cdot\widehat{n}_-;R] 
&= \int_0^\infty \dd r' \left(\frac{r'}{R}\right)^{d-2} \left(
\overline{G}_{d+1}^\text{(E)}\Big[ \vec{X}[R,\widehat{n}] - \vec{X}'[r',\widehat{n}_+] \Big] 
- \overline{G}_{d+1}^\text{(E)}\Big[ \vec{X}[R,\widehat{n}] - \vec{X}'[r',\widehat{n}_-] \Big]
\right) ,
\end{align}
with $\overline{G}_{d+1}^\text{(E)}$ in eq. \eqref{GreensFunction_Euclidean}, and the $X$ and $X'$ parametrization of eq. \eqref{dSphere_X_parametrization}. The integrals work out to yield, for $d \geq 2$,
{\allowdisplaybreaks\begin{align}
\label{dSphere_ModifiedGreensFunction_Even}
G_{\text{even }d}^\text{(E$\vert\pm$)}\left[ \widehat{n}\cdot\widehat{n}'_+, \widehat{n}\cdot\widehat{n}'_-; R \right] 
&= -\frac{\Gamma\left[ \frac{d-1}{2} \right]}{4\pi^{\frac{d+1}{2}} R^{d-2}} 
\left. \frac{(-)^{\frac{d-2}{2}}}{(1/2)_{\frac{d-2}{2}}} \left( \frac{\partial}{\partial a} \right)^{\frac{d-2}{2}} \left( \frac{1}{\sqrt{a}} \ln\left[ \frac{\sqrt{a} - \widehat{n} \cdot \widehat{n}'_-}{\sqrt{a} - \widehat{n} \cdot \widehat{n}'_+} \right] \right) \right\vert_{a=1} , \\
\label{dSphere_ModifiedGreensFunction_Odd}
G_{\text{odd }d}^\text{(E$\vert\pm$)}\left[ \widehat{n}\cdot\widehat{n}'_+, \widehat{n}\cdot\widehat{n}'_-; R \right] 
&= -\left.\frac{\Gamma\left[ \frac{d-1}{2} \right]}{4\pi^{\frac{d+1}{2}} R^{d-2}} 
\frac{(-)^{\frac{d-3}{2}}}{(1)_{\frac{d-3}{2}}}\left( \frac{\partial}{\partial a} \right)^{\frac{d-3}{2}} \right\vert_{a=1} \nonumber\\
&\qquad\qquad
\times \Bigg( 
	\frac{\widehat{n}\cdot\widehat{n}'_+}{a \sqrt{a - \left(\widehat{n} \cdot \widehat{n}'_+\right)^2}} \left(\frac{\pi}{2} + \tan^{-1}\left[\frac{\widehat{n} \cdot \widehat{n}'_+}{\sqrt{a - \left(\widehat{n} \cdot \widehat{n}'_+\right)^2}}\right]\right) \\
&\qquad\qquad\qquad\qquad
	- \frac{\widehat{n}\cdot\widehat{n}'_-}{a \sqrt{a - \left(\widehat{n} \cdot \widehat{n}'_-\right)^2}} \left(\frac{\pi}{2} + \tan^{-1}\left[\frac{\widehat{n} \cdot \widehat{n}'_-}{\sqrt{a - \left(\widehat{n} \cdot \widehat{n}'_-\right)^2}}\right]\right)
	\Bigg) , \nonumber
\end{align}}
A more symmetric version of these proposals, in that there is only one independent source vector $\widehat{n}'$ and that the Green's function is invariant under the exchange $\widehat{n} \leftrightarrow \widehat{n}'$, is achieved by setting $\widehat{n}_+ = -\widehat{n}_- \equiv \widehat{n}'$. (This means the positive and negative charges now lie on each other's antipodal points.)
{\allowdisplaybreaks\begin{align}
\label{dSphere_ModifiedGreensFunction_SymmetricVersion_Even}
G_{\text{even }d}^\text{(E$\vert$S)}\left[ \widehat{n}\cdot\widehat{n}'; R \right] 
&= -\frac{\Gamma\left[ \frac{d-1}{2} \right]}{4\pi^{\frac{d+1}{2}} R^{d-2}} 
\left. \frac{(-)^{\frac{d-2}{2}}}{(1/2)_{\frac{d-2}{2}}} \left( \frac{\partial}{\partial a} \right)^{\frac{d-2}{2}} \left( \frac{1}{\sqrt{a}} \ln\left[ \frac{\sqrt{a} + \widehat{n} \cdot \widehat{n}'}{\sqrt{a} - \widehat{n} \cdot \widehat{n}'} \right] \right) \right\vert_{a=1} , \\
\label{dSphere_ModifiedGreensFunction_SymmetricVersion_Odd}
G_{\text{odd }d}^\text{(E$\vert$S)}\left[ \widehat{n}\cdot\widehat{n}'; R \right] 
&= -\left.\frac{\Gamma\left[ \frac{d-1}{2} \right]}{4\pi^{\frac{d-1}{2}} R^{d-2}} 
\frac{(-)^{\frac{d-3}{2}}}{(1)_{\frac{d-3}{2}}}\left( \frac{\partial}{\partial a} \right)^{\frac{d-3}{2}} \left( \frac{\widehat{n}\cdot\widehat{n}'}{a \sqrt{a - \left(\widehat{n} \cdot \widehat{n}'\right)^2}} \right) \right\vert_{a=1} .
\end{align}}

{\bf Checks} \qquad By direct differentiation, we will now check that $\overline{\Box} G_d^\text{(E$\vert\pm$)} = 0$ almost everywhere. Because the results in equations \eqref{dSphere_ModifiedGreensFunction_Even} and \eqref{dSphere_ModifiedGreensFunction_Odd} can be expressed as a difference of two identical expressions, except one is a function of the sole variable $\cos\theta_+ \equiv \widehat{n}\cdot\widehat{n}_+$ and the other of the sole variable $\cos\theta_- \equiv \widehat{n}\cdot\widehat{n}_-$ \cite{CheckNoteI}, this means we may instead check that $\overline{\Box}$ acting on the portion of $G_d^\text{(E$\vert\pm$)}$ that depends only on $\cos\theta_+$ (or only on $\cos\theta_-$) gives a non-zero pure number \cite{CheckNoteII}. Now, the Laplacian $\overline{\Box}$ acting on a scalar function $\psi$ that depends on the angular coordinates $\vec{\theta}$ only through the dot product $\cos\theta_\pm$ is
\begin{align}
\label{dSphere_ReducedLaplacian}
\overline{\Box}_{\widehat{n}} \psi\left[ \widehat{n}[\vec{\theta}]\cdot\widehat{n}_\pm \right] 
= \frac{1}{\sin^{d-1}[\theta_\pm]} 
\frac{\partial}{\partial \theta_\pm} \left( \sin^{d-1}[\theta_\pm] \frac{\partial}{\partial \theta_\pm} \psi[ \cos\theta_\pm ] \right) .
\end{align}
Since the reduced Laplacian in eq. \eqref{dSphere_ReducedLaplacian} now depends only on $d$ and on a single angle $\theta_\pm$, it can be implemented readily on a computer \cite{Mathematica} -- we have ran this check for $d=2$ through $40$.

Let us further specialize to $d=2,3$, 
{\allowdisplaybreaks\begin{align}
\label{dSphere_ModifiedGreensFunction_d=2}
G_2^\text{(E$\vert\pm$)}\left[ \widehat{n}\cdot\widehat{n}'_+, \widehat{n}\cdot\widehat{n}'_-; R \right] 
&= -\frac{1}{4\pi} \left(
\ln\left[ 1 - \cos\theta_- \right] - \ln\left[ 1 - \cos\theta_+ \right]
\right) , \\
\label{dSphere_ModifiedGreensFunction_d=3}
G_3^\text{(E$\vert\pm$)}\left[ \widehat{n}\cdot\widehat{n}'_+, \widehat{n}\cdot\widehat{n}'_-; R \right] 
&= -\frac{1}{4\pi^2 R} \left(
\left( \pi - \theta_+ \right) \cot\theta_+ 
- \left( \pi - \theta_- \right) \cot\theta_- 
\right) ,
\end{align}}
and observe that the generalized Green's functions in equations \eqref{dSphere_ModifiedGreensFunction_d=2} and \eqref{dSphere_ModifiedGreensFunction_d=3} are regular everywhere except when the observer is right on top of one of the two sources at $\widehat{n}_\pm$. For $d=2$, this follows from the fact that the logs in \eqref{dSphere_ModifiedGreensFunction_d=2} are finite unless one of the cosines is unity. For $d=3$, $\cot\theta_\pm$ blows up at $\theta_\pm = 0,\pi$ but because of the factor $(\pi-\theta_\pm)$, we have $\lim_{\theta_\pm \to \pi} (\pi-\theta_\pm) \cot\theta_\pm = -1$. 

That there are only two points on the sphere where the field is unbounded, at least for $d=2,3$, supports our claim that equations \eqref{dSphere_ModifiedGreensFunction_Even} and \eqref{dSphere_ModifiedGreensFunction_Odd} are the fields generated by two (and only two) charges of opposite signs.

\section{Summary and Future Directions}
\label{Section_FutureDirections}

In this work, we have derived the formula in eq. \eqref{GreensFunction_GeneralFormula} to compute the minimally coupled massless scalar Green's function in a curved space(time) $g_{\mu\nu}[x] = P^2[y=y_0] \gb_{\mu\nu}[x]$ using the corresponding Green's function in flat space(time), if the former is embeddable in the latter via eq. \eqref{Metric_GeneralEmbeddingForm}. We have used it to work out the scalar Green's functions in de Sitter spacetime, and the results can be found in equations \eqref{GreensFunction_deSitter_IntegralRepresentation}, \eqref{GreensFunction_deSitter_Even}, \eqref{GreensFunction_deSitter_Odd}, and \eqref{GreensFunction_deSitter_4D_IofIII} through \eqref{GreensFunction_deSitter_4D_IIIofIII}. Instead of viewing the $d$-dimensional de Sitter scalar Green's function $G_d[x,x']$ as the field at $x$ produced by a point source at $x'$, our work has uncovered an alternate perspective: it is the field at $X[\rho=1/H,x]$ on the de Sitter hyperboloid produced by a line source in the ambient $(d+1)$D Minkowski spacetime, as parametrized in eq. \eqref{deSitter_X_parametrization}. The location where this line intersects the de Sitter hyperboloid, $X'[\rho'=1/H,x']$, is responsible for the scalar signal that propagates on the null cone (according to the de Sitter observer), and the rest of the same line is the source of the tail part of the scalar waves, which travels within the light cone of $x'$.

We then turned to the case of the Laplacian on the $d$-sphere, where we know no solution to the field generated by a single point source should exist because of topology. Equation \eqref{GreensFunction_GeneralFormula} broke down by acquiring a divergence that depended logarithmically on the cutoff introduced to regulate the answer, which in turn yielded a term that did not satisfy the zero mode equation \eqref{ZeroModeEquation}. The remedy we proposed was to solve for the field generated by one positive and one negative point charge, and this generalized Green's function for the $d$-sphere can be found in equations \eqref{dSphere_ModifiedGreensFunction_IntegralRepresentation} through \eqref{dSphere_ModifiedGreensFunction_SymmetricVersion_Odd}.

We now close by touching on a number of directions we wish to pursue further.

It should be possible to extend our arguments in section \eqref{Section_Generalities} to solve the Green's function of $\Box + m^2 + \xi \mathcal{R}$ in constant curvature geometries such as de Sitter. ($\mathcal{R}$ is the Ricci scalar and $\xi$ is a constant.) Also, the topology of the $d$-sphere is no longer an obstacle once a non-zero mass term is introduced: the charge density $J \equiv (\Box + m^2)\varphi$ does not have to integrate to zero. It may be instructive to solve the Green's function of $\Box+m^2$ on a sphere using the embedding method, and then take the $m \to 0$ limit to study exactly what goes bad.

In section \eqref{Section_deSitter} we discussed how the notion of causal influence in the $(d+1)$D Minkowski, as encoded in the integral representation of eq. \eqref{GreensFunction_deSitter_IntegralRepresentation}, is compatible with that on the $d$-dimensional de Sitter hyperboloid itself. It would deepen our understanding of the embedding paradigm and its range of applicability, if this discussion on causal influence can be made in the abstract, or, at the very least, examined for enough examples that general lessons can be extracted. It is also of physical significance to understand whether eq. \eqref{GreensFunction_GeneralFormula} can be extended to the case of higher rank tensors, including that of fermions. An important case is that of the linearized wave equations resulting from perturbing Einstein's field equations about some fixed geometry. (Note that it is often technically advantageous to first perform a scalar-vector-tensor decomposition of these equations before proceeding.) 

Finally, we wonder if there are other types of embeddings than the one in eq. \eqref{Metric_GeneralEmbeddingForm} that would allow us to solve for the curved space(time) Green's functions from flat ones. For instance, in \cite{Chu:2011ip} (building upon earlier work in \cite{DeWittDeWitt:1964}-\cite{PfenningPoisson:2000zf}), a perturbation theory was devised to solve for the scalar, vector and tensor Green's functions in the metric $g_{\mu\nu}^{(d)} \equiv \eta_{\mu\nu} + h_{\mu\nu}$, using the corresponding Green's functions in $\eta_{\mu\nu}$. On the other hand, we have seen in section \eqref{Section_Flat} that the $d$-dimensional flat spacetime Green's functions is sourced by a line extending into one higher spatial dimension, i.e., parallel to the $x^d$-axis. Does this embedding picture hold up once non-zero perturbations $h_{\mu\nu}$ are allowed? If so, is it possible to phrase the perturbation theory of \cite{Chu:2011ip} in terms of perturbation theory of how $g_{\mu\nu}^{(d)}$ can be embedded in $\eta_{\mathfrak{A}\mathfrak{B}}^{(d+1)}$, as well as what the line source would deform into?

\section{Acknowledgments}

I thank Lasha Berezhiani, Robert Brown, Dai De-Chang, Kurt Hinterbichler, Ted Jacobson, Austin Joyce, Justin Khoury, Denis Klevers, Hernan Piragua, and Dejan Stojkovic for discussions and questions. Much of the analytic calculations was done with {\sf Mathematica} \cite{Mathematica}. This work was supported by NSF PHY-1145525 and funds from the University of Pennsylvania.


\begin{thebibliography}{99}


\bibitem{Robertson:1933zz} 
  H.~P.~Robertson,
  Rev.\ Mod.\ Phys.\  {\bf 5}, 62 (1933).


\bibitem{Nomenclature}
{\it Nomenclature}: the portion of the field propagating inside the null cone is known in the literature as the tail.


\bibitem{Fronsdal:1959zza} 
  C.~Fronsdal,
  Phys.\ Rev.\  {\bf 116}, 778 (1959).

\bibitem{Kasner}
	E.~Kasner, Am.\ J.\ Math. {\bf 43}, 130 (1921)


\bibitem{Chu:2013xca} 
  Y.~-Z.~Chu,
  arXiv:1310.2939 [gr-qc].

\bibitem{NoteAdded}
{\it Note added}: After posting this paper and \cite{Chu:2013xca}, we found \cite{Bertola:2000mx}, which applied the embedding method to compute the Wightman functions of the scalar quantum field in de Sitter spacetime using their Minkowski counterparts. (See \S4.1 of \cite{Bertola:2000mx}.) In principle, these Wightman functions can be used to extract the retarded/advanced Green's functions and their causal structure. However, we emphasize that our main goal here is to show how one may instead arrive at them directly, so as to relate the causal structure of signals in the ambient Minkowski spacetime to that of the embedded hypersurface (e.g., de Sitter spacetime).

\bibitem{Bertola:2000mx} 
  M.~Bertola, J.~Bros, V.~Gorini, U.~Moschella and R.~Schaeffer,
  Nucl.\ Phys.\ B {\bf 581}, 575 (2000)
  [hep-th/0003098].
  

\bibitem{SoodakTiersten}
	H.~Soodak and M.~S.~Tiersten, Am. J. Phys. {\bf 61} (5), May 1993

\bibitem{GConfusion}
The $\bar{\mathcal{G}}$ here is not to be confused with the $\mathcal{G}$ in eq. \eqref{GreensFunction_GeneralFormula_Check_Integral}, though in the later sections we shall see that they are related.

\bibitem{G&S}
	I.~S.~Gradshteyn and I.~M.~Ryzhik, 
	``Table of Integrals, Series, and Products" 
	Edited by A. Jeffrey and D. Zwillinger, Academic Press, New York, 7th edition, 2007

\bibitem{FourierTransform}
The relations in equations \eqref{GreensFunction_IntegralRecursionBetween_d_and_d+1}, \eqref{GreensFunction_IntegralRecursionBetween_d_and_d+2}, and \eqref{GreensFunction_EuclideanFromMinkowski} can also be understood by replacing the Green's functions occurring within the integrals on the right hand side with their Fourier representations.

\bibitem{SourcesConfusion}
Which is not to be confused with the (related) line source in Minkowski -- at the risk of over-verbosity, we reiterate that the point source on the de Sitter hyperboloid is but one point on the line source in Minkowski.

\bibitem{CausalStructureAlgebra}
To understand these statements, note that the inverted parabola on the left hand side of eq. \eqref{deSitter_CausalCondition_MinkowskiSense} has no real roots, with respect to the variable $\rho'/\rho$, for $|Z|<1$. For $Z \leq -1$, the region where $(X-X')^2$ is positive lies between its two real roots. (The $Z > +1$ region gives negative $\rho'/\rho$, and is irrelevant for our purposes here.) Also observe that $\rho'/\rho = 0$, corresponding to one end of the line source, yields no solution to eq. \eqref{deSitter_CausalCondition_MinkowskiSense} because the light cone of $0^{\mathfrak{A}}$ in Minkowski does not intersect the hyperboloid of any finite size $\rho$.

\bibitem{RodriguesTail}
In eq. \eqref{GreensFunction_deSitter_Even_Tail}, we have used $\dd^m P_m[z]/\dd z^m = (2m)!/(2^m m!)$, for integers $m=0,1,2,3,\dots$; this follows from the Rodrigues formula for the Legendre polynomials. The tail of the de Sitter Green's function in even dimensions is therefore a constant.

\bibitem{Burko:2002ge} 
  L.~M.~Burko, A.~I.~Harte and E.~Poisson,
  Phys.\ Rev.\ D {\bf 65}, 124006 (2002)
  [gr-qc/0201020].

\bibitem{Poisson:2011nh} 
  E.~Poisson, A.~Pound and I.~Vega,
  Living Rev.\ Rel.\  {\bf 14}, 7 (2011)
  [arXiv:1102.0529 [gr-qc]].

\bibitem{Chu:2011ip} 
  Y.~-Z.~Chu and G.~D.~Starkman,
  Phys.\ Rev.\ D {\bf 84}, 124020 (2011)
  [arXiv:1108.1825 [astro-ph.CO]].

\bibitem{DeWittDeWitt:1964}
  C.~M.~DeWitt and B.~S.~DeWitt,
  Physics {\bf 1}, 3 (1964).

\bibitem{KovacsThorne:1975}
  S.~J.~Kovacs and K.~S.~Thorne,
  Astrophys.\ J.\  {\bf 200} (1975) 245.

\bibitem{PfenningPoisson:2000zf}
  M.~J.~Pfenning and E.~Poisson,
  Phys.\ Rev.\  D {\bf 65}, 084001 (2002)
  [arXiv:gr-qc/0012057].

\bibitem{PochhammerCalculation}
The result $\left. \frac{(-)^m}{(\beta)_m} \left( \partial_a^m \frac{1}{a^\beta} \right)\right\vert_{a=1} = 1$ for $\beta \neq 0$, is useful in evaluating some of the derivatives occurring in the intermediate steps.

\bibitem{CheckNoteI}
For even $d$, we may replace in eq. \eqref{dSphere_ModifiedGreensFunction_Even}, $\ln\left[ \frac{\sqrt{a} - \widehat{n} \cdot \widehat{n}'_-}{\sqrt{a} - \widehat{n} \cdot \widehat{n}'_+} \right] \to \ln\left[ \sqrt{a} - \cos\theta_- \right] - \ln\left[ \sqrt{a} - \cos\theta_+ \right]$.

\bibitem{CheckNoteII}
That it gives a non-zero pure number (as opposed to exactly zero) corroborates the topology based argument that we cannot interpret the $\cos\theta_+$ dependent portion of $G_d^\text{(E$\vert\pm$)}$ as the field generated by a single positive charge and the $\cos\theta_-$ dependent portion as that by a negative charge; it is only when their difference is taken, that $G_d^\text{(E$\vert\pm$)}$ is then annihilated by the Laplacian $\overline{\Box}$ almost everywhere.

\bibitem{Mathematica}
	Wolfram Research, Inc., Mathematica, Version 9.0.1.0, Champaign, IL (2013).
	
\end{thebibliography}
\end{document}